\documentclass[twocolumn]{svjour3}

\usepackage[hyperindex,breaklinks,colorlinks,citecolor=blue,urlcolor=blue]{hyperref}
\usepackage{mathtools, cuted}
\usepackage{lipsum}
\usepackage[longnamesfirst]{natbib}
\usepackage{float}
\usepackage{mathrsfs}
\usepackage{amsmath}
\numberwithin{equation}{section}
\usepackage{amssymb}
\usepackage[pdftex]{graphicx}
\usepackage{caption}
\usepackage{bbm}
\usepackage{mathrsfs}
\usepackage{amsfonts}
\usepackage{bm}
\usepackage{mathrsfs}
\usepackage{enumerate}
\usepackage[all]{xy}

\usepackage{xcolor}

\makeatletter

\newcommand{\m}[1]{\boldsymbol{#1}}

\newcommand{\B}{\mathcal{B}}
\newcommand{\D}{\bm{D}}

\newcommand{\M}{\mtx{M}}

\newcommand{\I}{\mathcal{I}}

\newcommand{\ol}{\overline}
\newcommand{\ul}{\underline}

\newcommand{\R}{\mathbb{R}}

\newcommand{\Z}{\mathbb{Z}}

\newcommand{\vct}[1]{\bm{#1}}
\newcommand{\mtx}[1]{\bm{#1}}

\newcommand{\grad}{\nabla}

\newcommand{\defby}{\mathrel{\mathop:}=}

\renewcommand{\l}{\left}
\renewcommand{\r}{\right}

\renewcommand{\phi}{\varphi}
\renewcommand{\epsilon}{\varepsilon}
\renewcommand{\H}{\mtx{H}}

\renewcommand{\v}{\vct{v}}
\newcommand{\w}{\vct{w}}

\newcommand{\h}{\vct{h}}
\newcommand{\mr}{\mathrm}

\newcommand{\x}{\vct{x}}

\renewcommand{\B}{\mtx{B}}

\newcommand{\y}{\vct{y}}
\newcommand{\yo}{\vct{y}^{\circ}}
\newcommand{\ho}{\ul{h}}

\newcommand{\U}{\vct{U}}

\newcommand{\ZZ}{\vct{Z}}
\renewcommand{\m}{\vct{m}}
\newcommand{\n}{\vct{n}}

\title{A 4D-Var Method with Flow-Dependent Background Covariances for the Shallow-Water Equations}

\author{Daniel Paulin \and Ajay Jasra \and Alexandros Beskos \and Dan Crisan}
\institute{Daniel Paulin
\at School of Mathematics, University of Edinburgh, James Clerk Maxwell Building, Peter Guthrie Tait Rd, Edinburgh, EH9 3FD, UK.\\\
\email{dpaulin@ed.ac.uk}
\and
Ajay Jasra
\at Computer, Electrical and Mathematical Science and Engineering Division, King Abdullah University of Science and Technology, Thuwal, 23955-6900, KSA.\\\
\email{ajay.jasra@kaust.edu.sa}
\and
Alexandros Beskos
\at Department of Statistical Science, University College London, London, WC1E 6BT, UK. \\\
\email{a.beskos@ucl.ac.uk}
\and
Dan Crisan
\at Department of Mathematics, Imperial College London, London, SW7 2AZ, UK.\\\
\email{d.crisan@ic.ac.uk}
}

\begin{document}

\maketitle

\begin{abstract}
The 4D-Var method for filtering partially observed nonlinear chaotic dynamical systems consists of finding the maximum a-posteriori (MAP) estimator of the initial condition of the system given observations over a time window, and propagating it forward to the current time via the model dynamics. This method forms the basis of most currently operational weather forecasting systems. In practice the optimization becomes infeasible if the time window is too long due to the non-convexity of the cost function, the effect of model errors, and the limited precision of the ODE solvers. Hence the window has to be kept sufficiently short, and the observations in the previous windows can be taken into account via a Gaussian background (prior) distribution. The choice of the background covariance matrix is an important question that has received much attention in the literature. In this paper, we define the background covariances in a principled manner, based on observations in the previous $b$ assimilation windows, for a parameter $b\ge 1$. The method is at most $b$ times more computationally expensive than using fixed background covariances, requires little tuning, and greatly improves the accuracy of 4D-Var.
As a concrete example, we focus on the shallow-water equations. The proposed method is compared against state-of-the-art approaches in data assimilation and is shown to perform  favourably on simulated data. We also illustrate our approach on data from the recent tsunami of 2011 in Fukushima, Japan.
\keywords{Filtering \and Smoothing \and Data Assimilation \and Gauss-Newton Method \and Shallow-Water Equations.}
\end{abstract}

\section{Introduction}

Filtering, or data assimilation, is a field of core importance in a wide variety of real applications, such as numerical weather forecasting, climate modelling and finance; see e.g. \cite{asch2016data, blayo2014advanced, Danmathofplanetearth, lahoz2010data, Dataassimilation} for an introduction. Informally, one is interested in carrying out inference about an unobserved signal process conditionally upon noisy observations. The type of unobserved process considered in this paper is that of a nonlinear chaotic dynamical system, with unknown initial condition. As an application in this paper we consider the case where the unobserved dynamics correspond to the discretised version of the shallow-water equations; see e.g.~\cite{salmontextbook}.
These latter equations are of great practical importance, generating realistic approximations of real world phenomena, useful in tsunami and flood modelling (see e.g. \cite{bates2010simple, pelinovsky2006hydrodynamics}).

For systems of this type, the filtering problem is notoriously challenging. Firstly, the filter is seldom available in analytic form due to the non-linearity. Secondly, even if the given system were solvable, the associated
dimension of the object to be filtered is very high (of order of $10^8$ or greater) thus posing great computational challenges.

One of the most successful methods  capable of handling such high dimensional datasets is the so-called 4D-Var algorithm \cite{le1986variational, talagrand1987variational}: it consists of optimizing a loss-functional so that under Gaussian noise it is equivalent to finding the maximum a-posteriori (MAP) estimator of the initial condition. Since its introduction, a lot of further developments in the 4D-Var methodology have appeared in the literature; for an overview of some recent advances, we refer the reader to \cite{bannister2016review, lorenc2014four, navon2009data,  park2009data, park2013data, park2017data}. The main focus of this article is to consider principled improvements of the 4D-Var algorithm.

%

An important practical issue of the 4D-Var method is that, due to chaotic nature of the systems encountered in weather prediction, the negative log-likelihood (cost function) can become highly non-convex if the assimilation window is too long. The reason for this is that for deterministic dynamical systems, as the assimilation window grows, the smoothing distribution gets more and more concentrated on the stable manifold of the initial position, which is a complicated lower dimensional set (see \cite{pires1996extending}, \cite{paulin2016concentration} for more details). On one hand, this means that it becomes very difficult to find the MAP estimator. On the other hand, due to the highly non-Gaussian nature of the posterior in this setting, the MAP might be far away from the posterior mean, and have a large mean square error. Moreover, for longer windows, the precision of the tangent linear model/adjoint solvers might decrease. Due to these facts, the performance of 4D-Var deteriorates for many models when the observation window becomes too long (see \cite{kalnay20074}).

The observations in the previous window are taken into account via the background (prior) distribution, which is a Gaussian whose mean is the estimate of the current position based on the previous windows, and has a certain covariance matrix.  The choice of this background covariance matrix is an important and difficult problem that has attracted much research. \cite{fisher2003background} states that in operational weather forecasting systems up to 85\% of the information in the smoothing distribution comes from the background (prior) distribution. The main contribution of this paper is an improvement of the 4D-Var methodology by a principled definition of this matrix in a flow-dependent way.  This is based on the observations in the previous $b$ assimilation windows (for a parameter $b\ge 1$).  Via simulations on the shallow-water model, we show that our method compares favourably in precision with the state-of-the-art Hybrid 4D-Var method (see \cite{lorenc2015comparison}).

The structure of the paper is as follows. In the rest of this section, we briefly review the literature on 4D-Var background covariances. In Section \ref{sec:model} the modelling framework for the shallow-water equations is described in detail.  In Section \ref{secflowdependent4DVAR} we introduce our 4D-Var method with flow-dependent background covariance. In particular, Section \ref{sec:comparisonwithflowdep} compares our method with other choices of flow-dependent background covariances in the literature. In Section \ref{secsimulations} we present some simulation results and compare the performance of our method with Hybrid 4D-Var. Finally, in Section \ref{secconclusion} we state some conclusions for this paper.

\subsection{Comparison with the literature}\label{sec:comparisonwithflowdep}
There exist mathematically rigorous techniques to obtain the filter with precision and use the mean of the posterior distribution as the estimate, based upon sequential Monte Carlo methods (e.g.~\cite{DDJJRRSB, rebeschini2015can}) which can provably work in high-dimensional systems \cite{BCJHD}. While these approximate the posterior means and hence are optimal in mean square error, and are possibly considerably more accurate than optimization based methods, nonetheless  such methodology can be practically overly expensive. As a result, one may have to resort to less accurate but more computationally efficient methodologies (see \cite{Dataassimilation} for a review). There are some relatively recent applications of particle filtering methods to high dimensional data assimilation problems, see e.g.~\cite{van2009particle, van2010nonlinear}. While these algorithms seem to be promising for certain highly non-linear problems, their theoretical understanding is limited at the moment due to the bias they introduce via various approximations. 

Despite the difficulty of solving the non-linear filtering problem exactly, due to the practical interest in weather prediction, several techniques have been devised and implemented operationally in weather forecasting centers worldwide. These techniques are based on optimization methods, hence they scale well to high dimensions and are able to process massive datasets. Although initially such methods were lacking in mathematical foundation, the books \cite{bengtsson1981dynamic} and \cite{kalnay} are among the first to open up the field of data assimilation to mathematics. Among the earlier works devoted to the efforts of bringing together data assimilation and mathematics, we also mention \cite{ghil1981applications} and \cite{ghil1991data}, where a comparison between the Kalman filter (sequential-estimation) and variational methods is presented.

The performance of 4D-Var methods depends very strongly on the choice of background covariances. One of the first principled ways of choosing 4D-Var background covariances was introduced by
\cite{parrish1992national}. They have proposed the so-called NMC method for choosing climatological prior error covariances based on a comparison of 24 and 48 hour forecast differences. This method was refined in \cite{derber1999reformulation}. \cite{fisher2003background} proposed the use of wavelets for forming background covariances; these retain the computational advantages of spectral methods, while also allow for spatial inhomogeneity. The background covariances are made flow-dependent via a suitable modification of the NMC approach. \cite{lorenc2003modelling} reviews some of the practical aspects of modelling 4D-Var error covariances, while \cite{fisher2005equivalence} makes a comparison between 4D-Var for long assimilation windows and the Extended Kalman Filter. As we have noted previously, long windows are not always applicable due to the presence of model errors and the non-convexity of the likelihood.

More recently, there have been several methods pro\-posing the use of ensembles combined with localization methods for modelling the covariances, see e.g. \cite{zupanski2005maximum, auligne2016ensemble, bannister2016review, bannister2008review1, bannister2008review2, bonavita2016evolution, bousserez2015improved,  buehner2005ensemble, clayton2013operational, hamill2011predictions, kuhl2013comparison, wang2013gsi}. Currently most operational NWP centers use the Hybrid 4D-Var method, which is based on a linear combination of a fixed background covariance matrix (the climatological background error covariance) and an ensemble based background covariance (see \cite{lorenc2015comparison}, \cite{fairbairn2014comparison}).

Localization eliminates spurious correlations betwe\-en elements of the covariance matrix that are far away from each other and hence they have little correlation. This means that these long range correlations are set to zero, which allows the sparse storage of the covariance matrix and efficient computations of the matrix-vector products. Bishop et al 2011 proposes an efficient implementation of localization by introducing some further approximations, using the product structure of the grid, and doing part of the calculations on lower resolution grid points. Such efficient implementations have allowed localized ENKF based background covariance modelling to provide the state-of-the-art performance in data assimilation, and they form the core of most operational NWP systems at the moment.

Over longer time periods, given sufficient data available, most of the variables become correlated, and imposing a localized structure over them leads to some loss of information vs the benefit of computational efficiency.  Our method does not impose such a structure as it writes the precision matrix in a factorized form. Moreover, the localization structure is assumed to be fixed in time, so even with a considerable amount of tuning for a certain time period of data it is not guaranteed that the same localization structure will be optimal in the future. Our method does not make such a constant localization assumption and hence it is able to adapt to different correlation structures automatically.

We use an implicit factorised form of the Hessian and the background precision matrix described in Sections \ref{secgradientHess}--\ref{secflowdependent4DVARfilter}, thus we only need to store the positions of the system started from the 4D-Var estimate of the previous $b$ windows at the observation times. This allows us to compute the effect of these matrices on a vector efficiently, without needing to store all the elements of the background precision matrix, which would require too much memory.

Although in this paper we have assumed that the model is perfect, there have been efforts to account for model error in the literature, see \cite{tr2006accounting}. The effect of nonlinearities in the dynamics and the observations can be in some cases so strong that the Gaussian approximations are no longer reasonable, see \cite{miller1994advanced, bocquet2010beyond, gejadze2011computation} for some examples and suggestions for overcoming these problems.

\section{Notations and Model}\label{sec:model}
\subsection{Notations}\label{subsec:notations}
In this paper, we will be generally using the unified notations for data assimilation introduced in \cite{ide1997unified}. In this section we briefly review the required notations for the 4D-Var data assimilation method.

The state vector at time $t$ will be denoted by $\x(t)$, and it is assumed that it has dimension $n$. The evolution of the system from time $s$ to time $t$ will be governed by the equation
\[\x(t)=M(t,s)[\x(s)],\]
where $M(t,s)$ is the model evolution operator from time $s$ to time $t$. In practice, this finite dimensional model is usually obtained by discretisation of the full partial differential equations governing the flow of the system. 

Observations are made at times $(t_i)_{i\ge 0}$, and they are of the form
\begin{equation}\label{yodefeq}
\yo_i=H_i[\x(t_i)]+\epsilon_i,
\end{equation}
where $H_i$ is the observation operator, and $\epsilon_i$ is the random noise.
We will denote the dimension $\yo_i$ by $n_{i}^{\circ}$, and assume that $(\epsilon_i)_{i\ge 0}$ are independent normally distributed random vectors with mean 0 and covariance matrix $(\mtx{R}_i)_{i\ge 0}$. The Jacobian matrix (i.e. linearization) of the operator $M(t,s)$ at position $\x(s)$ will be denoted by $\mtx{M}(t,s)$, and the Jacobian of $H_i$ at $\x(t_i)$ will be denoted by $\mtx{H}_i$. The inverse and transpose of a matrix will be denoted by $(\cdot)^{-1}$ and $(\cdot)^{T}$, respectively.

The 4D-Var method for assimilating the observations in the time interval $[t_0,t_{k-1}]$ consists of minimizing the cost functional 
\begin{align}
\nonumber J[\x(t_0)]&=\frac{1}{2}[\x(t_0)-\x^{b}(t_0)]^{T}\B_0^{-1} [\x(t_0)-\x^{b}(t_0)]\\
&+\frac{1}{2}\sum_{i=0}^{k-1}[\y_i-\yo_i]^{T}\mtx{R}_i^{-1} [\y_i-\yo_i],
\label{Jdefeq}
\end{align}
where $\y_i\defby H_i(\x(t_i))$, and $\B_0$ denotes the background covariance matrix, and $\x^{b}(t_0)$ denotes the background mean. Minimizing this functional is equivalent to maximizing the likelihood of the smoothing distribution for $\x(t_0)$ given $\yo_{0:{k-1}}:=\{\yo_{0},\ldots, \yo_{k-1}\}$ and normally distributed prior with mean $\x^{b}(t_0)$ and covariance $\B_0$.
Note that the cost function \eqref{Jdefeq} corresponds to the so-called \emph{strong constraint} 4D-Var (i.e. no noise is allowed in the dynamics), there are also \emph{weak constraint} alternatives that account for possible model errors by allowing noise in the dynamics (see e.g.~\cite{tr2006accounting}).

\subsection{The Model}\label{subsec:model}

We consider the shallow-water equations, e.g.~as described in \cite[pg.~105-106]{salmontextbook}, but with added diffusion and bottom friction terms, i.e.
\begin{align}
\frac{\partial u}{\partial t}&=\l(-\frac{\partial u}{\partial y}+f\r) v-\frac{\partial}{\partial x}\l(\frac{1}{2}u^2+gh\r)+\nu \grad^2 u -c_b u; \\
\frac{\partial v}{\partial t}&=-\l(\frac{\partial v}{\partial x}+f\r) u-\frac{\partial}{\partial y}\l(\frac{1}{2}v^2+gh\r)+\nu \grad^2 v -c_b v;\\
\frac{\partial h}{\partial t}&=-\frac{\partial}{\partial x}((h+\ho)u) -\frac{\partial}{\partial y}((h+\ho)v).
\end{align}
Here, $u$ and $v$ are the velocity fields in the $x$ and $y$ directions respectively, and $h$ the
field for the height of the wave. Also,  $\ho$ is the depth of the ocean, $g$ the gravity constant,  $f$   the Coriolis parameter, $c_b$  the bottom friction coefficient and $\nu$  the viscosity coefficient. Parameters $\ho$, $f$, $c_b$ and $\nu$ are assumed to be constant in time but in general depend on the location. The total height of the water column is the sum $\ul{h}+h$. 

For square grids, under periodic boundary conditions, the equations are discretised as 
\begin{align}
&\nonumber\frac{d u_{i,j}}{dt} =f_{i,j} v_{i,j}-\frac{g}{2\Delta}(h_{i+1,j}-h_{i-1,j})\\
\nonumber& -c_b u_{i,j}+\frac{\nu}{\Delta^2} \l(u_{i+1,j}+u_{i-1,j}+u_{i,j+1}+u_{i,j-1}-4u_{i,j}\r)\\
\label{ueq}&  -\frac{1}{2\Delta} \l[\l(u_{i,j+1}-u_{i,j-1}\r)v_{i,j} + \l(u_{i+1,j}-u_{i-1,j}\r)u_{i,j}\r],
\\[0.2cm]
\nonumber &\frac{d v_{i,j}}{dt} =-f_{i,j} u_{i,j}-\frac{g}{2\Delta} (h_{i,j+1}-h_{i,j-1})\\
\nonumber& -c_b v_{i,j}+\frac{\nu}{\Delta^2} \l(v_{i+1,j}+v_{i-1,j}+v_{i,j+1}+v_{i,j-1}-4v_{i,j}\r)\\
\label{veq}& -\frac{1}{2\Delta} \l[\l(v_{i+1,j}-v_{i-1,j}\r)u_{i,j} +\l(v_{i,j+1}-v_{i,j-1}\r)v_{i,j}\r],
\end{align}
\begin{align}
\nonumber&\frac{d h_{i,j}}{dt} =\\
\nonumber&-\frac{1}{2\Delta}\l(h_{i,j}+\ho_{i,j}\r)\l(u_{i+1,j}-u_{i-1,j}+v_{i,j+1}-v_{i,j-1}\r)\\ &
\nonumber -\frac{1}{2\Delta}u_{i,j}\l(h_{i+1,j}+\ho_{i+1,j}-h_{i-1,j}-\ho_{i-1,j}\r)\\
&\label{heq} -\frac{1}{2\Delta}v_{i,j}\l(h_{i,j+1}+\ho_{i,j+1}-h_{i,j-1}-\ho_{i,j-1}\r),
\end{align}
where $1\le i,j\le d$, for a typically large $d\in \mathbb{Z}_{+}$, with the indices understood modulo~$d$ (hence the domain is a torus), and some space-step $\Delta>0$. Summing up \eqref{heq} over $1\le i,j\le d$, one can see that the discretisation preserves the total mass $h_{\mr{tot}}\defby \sum_{i,j}(h_{i,j}+\ho_{i,j})$. If we assume that the viscosity and   bottom friction are negligible, i.e.~$\nu=c_b=0$, then the total energy \begin{align*}
&E_{\mr{tot}}\defby  \frac{1}{2}\cdot\\
& \sum_{i,j} \l((h_{i,j}+\ho_{i,j})u_{i,j}^2 + (h_{i,j}+\ho_{i,j})v_{i,j}^2+g(h_{i,j}^2-\ho_{i,j}^2)\r)
\end{align*}
is also preserved. When the coefficients $c_b$ and $\nu$ are not zero, the bottom friction term always decreases the total energy (the sum of the kinetic and potential energy), while the diffusion term tends to smooth the velocity profile. We denote the solution of  equations \eqref{ueq}-\eqref{heq} at time $t\ge 0$ as
\begin{align*}
&\x(t)\defby\\
&\left(\l(u_{i,j}(t)\r)_{1\le i,j\le d}, \l(v_{i,j}(t)\r)_{1\le i,j\le d}, \l(h_{i,j}(t)\r)_{1\le i,j\le d}\r).
\end{align*}
The unknown and random initial condition is denoted by $\x(0)$. One can show by standard methods (see \cite{murray2013existence}) that the solution of \eqref{ueq}-\eqref{heq} exists up to some time $T_{sol}(\x(0))>0$. In order to represent the components of $\x(t)$, we introduce a vector index notation. The set $\I\defby \{u,v,h\}\times \{1,\ldots,d\}\times \{1,\ldots,d\}$ denotes the possible indices, with the first component referring to one of $u$, $v$, $h$, the second component   to coordinate $i$, and the third  to   $j$. A vector index in $\I$ will usually be denoted as $\m$ or $\n$, e.g.\@ if $\m=(u,1,2)$, then $\x_{\m}(t)\defby  u_{1,2}(t)$. 

We assume that the $n\defby 3d^2$ dimensional system is observed at time points $(t_l)_{l\ge 0}$, with observations described as in Section \ref{subsec:notations}.
The aim of smoothing and filtering is to approximately reconstruct $\x(t_0)$ and $\x(t_{k})$ based on observations $\yo_{0:{k-1}}$. We note that data assimilation for the shallow-water equations have been widely studied in the literature, see \cite{bengtsson1981dynamic}, \cite{egbert1994topex} for the linearised form and \cite{lyard2006modelling}, \cite{courtier1990variational},  \cite{ghil1991data} for the full non-linear form of the equations.

\section{4D-Var with Flow-Dependent Covariance}\label{secflowdependent4DVAR}
\subsection{Method Overview}\label{secflowdependent4DVARoverview}
Assume that observations $\yo_{0:k-1}$ are made at time po\-ints $t_l=t_0+l h$ for $l=0,\ldots, k-1$, and let $T\defby kh$. 
The 4D-Var method for assimilating the observations in the time interval $[t_0,t_{k-1}]$ consists of minimizing the cost functional \eqref{Jdefeq}.
Under the independent Gaussian observation error assumption, $-J[\x(t_0)]$ is the log-likelihood of the smoothing distribution, ignoring the normalising constant. The minimizer of $J$ is the MAP estimator, and is denoted by $\hat{\x}_0$ (if multiple such minimizers exist, then we choose any of them). A careful choice of the background distribution is essential, especially in the case when the total number of observations in the assimilation window is smaller than the dimension of the dynamical system, where without the prior distribution, the likelihood would be singular (see   \cite{stuartbayesianapproach} for a principled method of choosing priors).


To obtain the MAP estimator,  we make use of Newton's method. Starting from some appropriate initial position $\x_0\in \R^{n}$, the method proceeds via the iterations
\begin{equation}\label{Newtoneq}
\x_{l+1}=\x_l-\l(\frac{\partial^2 J}{\partial \x_l^2}\r)^{-1} \frac{\partial J}{\partial \x_l},\quad l\ge 0,
\end{equation}
where $\frac{\partial J}{\partial \x_l}$ and $\frac{\partial^2 J}{\partial \x_l^2}$ denote the gradient and Hessian of $J$ at $\x_l$, respectively. Due to the high dimensionality of the systems in weather forecasting, typically iterative methods such as the preconditioned conjugate gradient (PCG) are used for evaluating \eqref{Newtoneq}. The iterations are continued until the step size $\|\x_{l+1}-\x_{l}\|$ falls below a   threshold $\delta_{\min}>0$.  The final position is denoted by $\hat{\x}_{*}$, and this is the numerical estimate for $\hat{\x}_0$ - with its push-forward $M(t_k,t_0)[\hat{\x}_{*}]$ then being  the numerical estimate for $M(t_k,t_0)[\hat{\x}_{0}]$.

To apply the iterations \eqref{Newtoneq}, one needs to compute the gradient and the Hessian of $J$ (or, more precisely, the application of the latter to a vector, which is all that is required for iterative methods such as PCG). An efficient method for doing this is given in the next section.  In practice, one cannot apply the above optimization procedure for arbitrarily large $k$ due to the non-convexity of the smoothing distribution for big enough $k$ (due of the nonlinearity of the system). Therefore, we need to partition the observations into blocks of size $k$ for some reasonably small $k$, and apply the procedure on them separately. The observations in the previous blocks can be taken into account by appropriately updating the prior distribution. The details of this procedure are explained in Section \ref{secflowdependent4DVARfilter}. 
Finally, in Section \ref{sec:comparisonwithflowdep} we compare our method with other choices of flow-dependent background covariances in the literature.

\subsection{Gradient and Hessian Calculation}\label{secgradientHess}
We can rewrite the gradient and Hessian of the cost function $J$ at a point $\x(t_0)\in \R^n$ as
%
\begin{align}
\nonumber&\frac{\partial J}{\partial \x(t_0)}=\mtx{B}_0^{-1}[\x(t_0)-\x^{b}(t_0)]\\
\label{gradgsmeq}&-\sum_{l=0}^{k-1}\mtx{M}(t_{l},t_0)^{T}\H_l^{T} \mtx{R}_l^{-1} (\y_l-\yo_l),\\
&\nonumber\frac{\partial^2 J}{\partial \x(t_0)^2}=\mtx{B}_0^{-1}+\sum_{l=0}^{k-1}
\mtx{M}(t_{l},t_0)^{T}\H_l^{T} \mtx{R}_l^{-1}\H_l\mtx{M}(t_{l},t_0)\\
\nonumber&-\sum_{l=0}^{k-1}\l(\frac{\partial M(t_{l},t_0)}{\partial \x_0^2}\r)^{T}\H_l^{T} \mtx{R}_l^{-1} (\y_l-\yo_l)-\\
&\sum_{l=0}^{k-1}\l(\mtx{M}(t_{l},t_0)^{T}\r)^2\left(\frac{\partial^2 H_l}{\partial \x(t_l)^2}\right)^{T} \mtx{R}_l^{-1} (\y_l-\yo_l).
\label{Hessgsmeq}
\end{align}
%
%
%
Let $\M_{i}\defby \M(t_{i},t_{i-1})$, then $\M(t_{l},t_0)=\M_l\cdot \ldots \cdot \M_1$, so the sum in the gradient (\ref{gradgsmeq}) can be rewritten as
\begin{align}
\nonumber&\sum_{l=0}^{k-1}\mtx{M}(t_{l},t_0)^{T}\H_l^{T} \mtx{R}_l^{-1} (\y_l-\yo_l)\\
&=\sum_{l=0}^{k-1}\M_1^T\cdot \ldots \cdot \M_l^T \H_l^T \mtx{R}_l^{-1} (\y_l-\yo_l)
\label{eq:iden}
\end{align}
The above summation can be efficiently performed as follows.
We consider the sequence of vectors
%
%
\begin{gather*}
\vct{g}_{k-1}:= \H_{k-1}^T\mtx{R}_{k-1}^{-1}\l(\y_{k-1}-\yo_{k-1}\r); \\
\vct{g}_l:=\H_{l}^T \mtx{R}_l^{-1} (\y_l-\yo_l)+\M_{l+1}^T\vct{g}_{l+1}, \quad k-1>l\ge 0.
\end{gather*}
The sum on the right side of \eqref{eq:iden} then equals $\vct{g}_{0}$. We note that this  method of computing the gradients forms the basis of the \emph{adjoint} method,  introduced in \cite{talagrand1987variational}, see also \cite{talagrand1997assimilation}.

In the case of the Hessian, in \eqref{Hessgsmeq} there are also second order Jacobian terms. If $\x(t_0)$ is close to the true initial position, then $(\y_l-\yo_l)\approx \epsilon_l$. Therefore in the low-noise/high-frequency regime, given a sufficiently precise initial estimator, these second order terms can be neglected. Using such Hessian corresponds to the so-called Gauss--Newton method, which has been studied in the context of 4D-Var in \cite{Lawless_GaussNewton}.
Thus, we use the approximation
\begin{align}
\widehat{\frac{\partial^2 J}{\partial \x(t_0)^2}}:=\mtx{B}_0^{-1}+\sum_{l=0}^{k-1}
\mtx{M}(t_{l},t_0)^{T}\H_l^{T} \mtx{R}_l^{-1}\H_l\mtx{M}(t_{l},t_0) \label{gsmHessapprox1}
\end{align}
A practical advantage of removing the second order terms is that if the Hessian of the log-likelihood of the prior, $\mtx{B}_0$ is positive
definite, then the resulting sum is positive definite, so the direction of $-\left(\widehat{\frac{\partial^2 J}{\partial \x(t_0)^2}}\right)^{-1}\cdot \frac{\partial J}{\partial \x(t_0)}$ is always a direction of descent (which is not always true if the second order terms are included). Note that via the so-called second-order adjoint equations, it is possible to avoid this approximation, and compute the  action of the Hessian $\frac{\partial^2 J}{\partial \x(t_0)^2}$ on a vector in $O(n)$ time, see \cite{le2002second}. However this can be slightly more computationally expensive, and in our simulations the Gauss-Newton approximation \eqref{gsmHessapprox1} worked well.

For the first order terms in the Hessian, for any $\w\in \R^{n}$, we have
\begin{align}
\nonumber
&\sum_{l=0}^{k-1}
\mtx{M}(t_{l},t_0)^{T}\H_l^{T} \mtx{R}_l^{-1}\H_l\mtx{M}(t_{l},t_0) \w\\
&=\sum_{l=0}^{k-1}\M_1^T\M_2^T\ldots \M_l^T\H_l^{T} \mtx{R}_l^{-1}\H_l\M_l\M_{l-1}\ldots \M_1 \w.\label{hesssumeq1}
\end{align}
We define
\begin{equation*}
\w_l:=\M_l\ldots\M_1\w, \quad l=0,\ldots, k-1;
\end{equation*}
and consider the sequence of vectors
\begin{gather}
\vct{h}_{k-1}=\H_{k-1}^{T} \mtx{R}_{k-1}^{-1}\H_{k-1}\w_{k-1}; \nonumber \\
\h_l=\H_{l}^{T} \mtx{R}_l^{-1}\H_{l} \w_{l}+\M_{l+1}^T\h_{l+1},\quad k-1> l\ge 0.
\label{hesssumeq2}
\end{gather}
%
%
Then the sum on the right side of \eqref{hesssumeq1} equals $\h_0$. The Hessian plays an important role in practical implementations of the 4D-Var method, and several methods have been proposed for its calculation (see \cite{courtier1994strategy, le2002second, lawless2005investigation}). Due to computational considerations, usually some approximations such as lower resolution models are used when computing Hessian-vector products for Krylov subspace iterative solvers in practice (this is the so-called incremental 4D-Var method, see \cite{courtier1994strategy}). Note that it is also possible to use inner and outer loops, where in the inner loops both the Hessian-vector products and the gradient are run on lower resolution models, while in the outer loops we use the high resolution model for the gradient, and lower resolution model for the Hessian-vector products. \cite{lawless2005investigation} has studied the theoretical properties of this approximation. In practice, the speedup from this method can be  substantial, but this approximation can introduce some instability, hence appropriate tuning is needed to ensure good performance. 

At the end of Section \ref{secflowdependent4DVARfilter}, we discuss how can the incremental 4D-Var strategy be combined with the flow-dependent background covariances proposed in this paper.

\subsection{4D-Var Filtering with Flow-Dependent Covariance}\label{secflowdependent4DVARfilter}

In this section we describe a 4D-Var based filtering procedure that can be implemented in an online fashion, with observations $\{\yo_l\}_l$ obtained at times $t_l=lh$, $l=0,1,\ldots$ (although the method can be also easily adapted to the case when the time between the observations varies).
We first fix an assimilation window length $T=kh$, for some $k\in \Z_+$, giving rise to
consecutive windows $[0,t_k],\,[t_k,t_{2k}],\ldots$.

Let the background distribution on $\x(t_0)$ be Gaussian with mean   $\x^{b}(t_0)$ and covariance matrix $\B_0$. In general, let 
the background distributions for the position of the signal at the beginning of each assimilation window, $\{\x(t_{mk})\}_{m\ge 0}$, have means $\{\x^{b}(t_{mk})\}_{m\ge 0}$ and covariance matrices $\{\B_{mk}\}_{m\ge 0}$. There are several ways to define these quantities sequentially, as we shall explain  later on in this section. Assuming that these are determined with some approach, working on the window $[t_{mk},t_{(m+1)k}]$ we set our estimator $\hat{\x}({t_{mk}})$ of $\x(t_{mk})$ as the MAP of the posterior of $\x(t_{mk})$
given background with mean $\x^{b}(t_{mk})$ and covariance $\B_{mk}$, and data $\yo_{mk:(m+1)k-1}$; we also obtain estimates
for subsequent times in the window, via push-forward, i.e.
\[\hat{\x}(t_l):=M(t_{l},t_{mk})[\hat{\x}(t_{mk})]$, $mk\le l<(m+1)k.\]
Recall that the numerical value of MAP is obtained
by the Gauss--Newton method (see~\eqref{Newtoneq},
with the details given in Section \ref{secflowdependent4DVARoverview}).

We now discuss choices for the specification of the background distributions.
A first option is to set these distributions identical to the first one, and set $\B_{mk}\defby \B_{0}$ and $\x(t_{mk})\defby \x(t_0)$ (i.e.~no connection with earlier observations). A second choice  (used in the first practical implementations of the 4D-Var method) is to set $\B_{mk}\defby \B_{0}$ (the covariance is kept constant) but change the background mean to 
\begin{equation}\label{backgroundmeaneq}
\x^{b}(t_{mk}):=M(t_{mk},t_{(m-1)k})[\hat{\x}(t_{m(k-1)})],
\end{equation}
i.e.~adjusting the prior mean to earlier observations. Finally, one can attempt to update both the mean and the covariance matrix of the background (prior) distribution, and this is the approach we follow here.

Note that we still define the background means according to \eqref{backgroundmeaneq}. To obtain data-informed background covariances $\B_{mk}$ we use Gaussian approximations for a number, say $b\ge 1$, of earlier windows of length $T$, and appropriately push-forward these to reach the instance of current interest $t_{mk}$.
There are two reasons why we use a fixed $b$ and do not push-forward all the way from time $t_0$. The first is to avoid quadratic costs in time. The total computational cost for our approach up to time $mT$ scales linearly with time for a fixed $b$, but if we would start from $t_0$, then we would incur $O(m^2)$ computational cost (or if it is done by storing the whole covariance matrix directly, then the approach would have $O(d^2)$ memory cost which is prohibitive in practice). The second reason is that a Gaussian distribution propagated through non-linear dynamics for longer and longer intervals of length $bT$ becomes highly non-Gaussian for large values of $b$, so the resulting background distribution can lead to poorer results than using smaller values of $b$. Reminiscent to 4D-Var, at time $t_{(m-b)k}$ we always start off the procedure with the same background
covariance $\B_0$.
%
In \cite{optimization2017} it was shown -- under certain assumptions -- that for a class of non-linear dynamical systems, for a fixed observation window $T$, if  $\|\mtx{R}_i\|=O(\sigma^2)$ and $\sigma\sqrt{h}$ is sufficiently small ($h$ is the observation time step) then the smoothing and filtering distributions can indeed be well approximated by Gaussian laws.
Following the ideas behind (\ref{gsmHessapprox1}),
an approximation of the Hessian of $J$, evaluated at the MAP given data $\yo_{(m-1)k:mk-1}$ is given as 
\[
\mtx{B}_{(m-1)k}^{-1}+\mtx{D}_{(m-1)k:mk-1},\]

where we
have defined
\begin{gather*}\mtx{D}_{(m-1)k:mk-1}:=\sum_{l=0}^{k-1} \mtx{A}_{m-1,l}^T \mtx{R}_{(m-1)k+l}^{-1}\mtx{A}_{m-1,l};\\
\mtx{A}_{m-1,l}:=\\
\H_{(m-1)k+l}\mtx{M}(t_{(m-1)k+l},t_{(m-1)k})[\hat{\x}(t_{(m-1)k})]. \end{gather*}
If the precision (inverse covariance) of the background were 0, then $\mtx{D}_{(m-1)k:mk-1}$ would correspond to the Hessian of $J$ at the MAP, and the smoothing distribution could be approximated by a Gaussian with mean $\hat{\x}(t_{(m-1)k})$ and precision matrix $\mtx{D}_{(m-1)k:mk-1}$.
Recall the change of variables formula:
if $\ZZ\sim N(m, \mtx{P}^{-1})$ in $\R^{n}$ and $\phi:\R^{n}\to \R^{n}$ is a continuously differentiable function, then $\phi(\ZZ)$ follows approximately
\begin{equation}\label{normalpushforwardeq}
N\l(\phi(m), \l[\l(\l(\frac{\partial \phi}{\partial m} \r)^{-1}\r)^{T}\cdot \mtx{P} \cdot \left(\frac{\partial \phi}{\partial m}\r) ^{-1}\r]^{-1}\r).
\end{equation}
The quality of this approximation depends on the size of the variance of $\ZZ$, and the degree of non-linearity of $\phi$. A way to consider the effect of the observations in the previous $b$ assimilation windows is therefore by using the recursion
\begin{align}
&\B_{(m-b)k}^{m} = \B_0; \nonumber \\
 \nonumber &\B_{(m-b+j)k}^{m} \\
\nonumber &= \bigg[\l(\l(\M(t_{(m-b+j)k},t_{(m-b+j-1)k})[\hat{\x}(t_{(m-b+j-1)k}]\r)^{T}\r)^{-1}\\
 \nonumber&\cdot \bigg( \l(\B_{(m-b+j-1)k}^{m}\r)^{-1} + \D_{(m-b+j-1)k:(m-b+j)k-1} \bigg)\\
 &\cdot \l(\M(t_{(m-b+j)k},t_{(m-b+j-1)k})[\hat{\x}(t_{(m-b+j-1)k}]\r)^{-1}\bigg]^{-1}, 
\label{Qmkdef1}
\end{align}
for $j=1,\ldots, b$, and set $\B_{mk}:=\B_{mk}^{m}$, where we have defined
\begin{gather*}\D_{(m-b+j-1)k:(m-b+j)k-1}:=\\
\sum_{l=0}^{k-1} \mtx{A}_{m-b+j-1,l}^T \mtx{R}_{(m-b+j)k+l}^{-1} \mtx{A}_{m-b+j-1,l}, \quad j=1, \ldots, b;\\
\mtx{A}_{m-b+j-1,l}:=\H_{(m-b+j-1)k+l}\\
\cdot \mtx{M}(t_{(m-b+j-1)k+l},t_{(m-b+j-1)k})[\hat{\x}(t_{(m-b+j-1)k})].
\end{gather*}
Note that similarly to the idea of variance inflation for the Kalman filter, one could include a multiplication by an inflation factor $(1+\alpha)$ for some $\alpha>0$ in the definition of $\B_{(m-b)k}^{m}$ in \eqref{Qmkdef1}. To simplify the expressions \eqref{Qmkdef1}, we define
%
%
%
\begin{align}
\nonumber&\M_{-j}\defby \M(t_{(m-j+1)k},t_{(m-j)k}))[\hat{\x}(t_{(m-j)k})],\\
&j=1,2,\ldots,b.\label{Jmldefeq}
\end{align}
%
The action of $\B_{mk}^{-1}$ on a vector $\w\in \R^{n}$ can be computed efficiently as follows. Let
\begin{equation}
\label{wminuseq} \w_{-j}:=\M_{-j}^{-1}\cdots \M_{-1}^{-1}\w, \quad j=1,2,\ldots,b.
\end{equation}
We then determine the recursion
\begin{gather}
\mathcal{B}_{-b}:=(\M_{-b}^T)^{-1}(\B_0^{-1}+\D_{(m-b)k:(m-b+1)k-1})\w_{-b}\nonumber\\
\mathcal{B}_{-j}:=(\M_{-j}^T)^{-1}(\mathcal{B}_{-j+1}+\mtx{D}_{(m-j)k:(m-j+1)k-1}\w_{-j}),\nonumber\\
 j=b-1, \ldots, 1. \label{Qminuseq}
\end{gather}
%
Then it is easy to see that $\B_{mk}^{-1}\w=\mathcal{B}_{-1}$. 

In order to evaluate the quantities in   \eqref{wminuseq} and \eqref{Qminuseq} for the shallow-water equations \eqref{ueq}-\eqref{heq}, we need implement the 
effect of the Jacobians $\M_1,\ldots, \M_k$, their inverses $\M_1^{-1},\ldots, \M_k^{-1}$, and their transpose for the previous $b$ assimilation windows. Note that multiplying by $\mtx{D}_{(m-l)k:(m-l+1)k-1}$ is equivalent to evaluating $\eqref{hesssumeq2}$ for the appropriate Jacobians, hence it is also based on multiplication by these Jacobians, their inverses and their transposes.

Matrix-vector products of the form $\M_j \v$ and $\M_j' \v$ can be computed by the tangent linear model, and by the adjoint equations, respectively. When computing matrix-vector products of the form $\M_j^{-1}\v$ and $(\M_j^{-1})'\v$, we need to run the tangent linear model backward in time, while the adjoint equations forward in time. It is important to note that while normally this would lead to numerical instability if done for a long time period (as the shallow-water equations are dissipative), this is not a problem here as we only run them over short time periods, the time between two observations (even shorter time periods could be possible if needed by breaking the Jacobians into products of Jacobians over shorter intervals). The initial point of these backward runs of the original equation (and forward runs of the adjoint equation) is always based on a forward run of the original equation, hence the instability is avoided. For the shallow-water equations, the Jacobians $\M_j$ can be stored directly in a sparse format, see the Appendix for more details (this reduces the need to use the ODE solvers repeatedly during the optimization  steps, however, this is not necessary  for the method to work as we can always use the adjoint/tangent linear solvers directly as described above).


Since we use the forward and adjoint equations of the model, computational cost of using these $b$ previous intervals in the calculation of the gradient and the product of the Hessian with a vector is a Newton's step is at most $O(b)$ times more than just using the observations in the current window. 
The key idea behind the choice of the precision matrices \eqref{Qmkdef1} is that we approximate the likelihood terms corresponding to the observations in the previous windows by Gaussian distributions, and then propagate them forward to the current time position via the Jacobians of the dynamics according to the change of variables formula \eqref{normalpushforwardeq}. This allows us to effectively extend the assimilation time $T$ to $(b+1)T$, but without the non-convexity issue that would occur if it would be extended directly (this was confirmed in our simulations). Moreover, the choice \eqref{Qmkdef1} introduces a strong linkage between the successive assimilation windows, and effectively allows the smoothing distribution to rely on two sided information (both from the past and the future), versus one sided information if one would simply use a longer window of length $(b+1)T$.
In fact, this was confirmed during our simulations, and we have found that increasing $T$ beyond a certain range did not improve the performance, while increasing $b$ has resulted in an improvement in general up to a certain point. \\

%
We note that the incremental 4D-Var strategy of \cite{courtier1994strategy} can be implemented here as follows. In the inner loops, we compute the gradient $\frac{\partial J}{\partial \x(t_0)}$ using the the adjoint equations with lower resolution models (see \eqref{gradgsmeq}) . The Hessian-matrix products required to compute the $\mtx{B}_0^{-1}[\x(t_0)-\x^{b}(t_0)]$ term in the gradient from the flow-dependent matrix covariances can also be computed using lower resolution models. For the Hessian matrix products in the iterative Krylov subspace solvers, we can always use the lower resolution model.

In contrast with this, in the outer loops, when computing the gradient $\frac{\partial J}{\partial \x(t_0)}$ we always need to use the highest resolution model (including in Hessian-matrix products required to compute the $\mtx{B}_0^{-1}[\x(t_0)-\x^{b}(t_0)]$ term in the gradient). The Hessian-matrix products for the iterative solvers can still be computed on lower resolution models.

\section{Simulations}\label{secsimulations}
In this section, we are going to illustrate the performance of our proposed method through simulation results. As a comparison, we also apply the Hybrid 4D-Var method on the same datasets as these form the basis of most currently used data-assimilation systems (see \cite{clayton2013operational}, \cite{lorenc2015comparison}). 
Section 5 of \cite{kalnay}, \cite{evensen2009data} and Sections 7-8 of \cite{reichcotterbook} offer excellent introductions to standard data assimilation methods such as 4D-Var and ENKF and its variants.

We consider two linear observation scenarios. In both of them, it is assumed that the observations happen in every $h$ time units, and that the linear observation operators $H_i$ are the same each time, represented by a matrix $\mtx{H}\in \R^{n^{\circ}\times n}$. The scenarios are as follows.

\begin{enumerate}
	\item[1.]{We observe the height $h$  at every gridpoint $1\le i,j\le d$, and the  velocities $u$ and $v$  at selected locations with spatial frequency $r$ in both directions for a positive integer $r$. All of the observation errors are i.i.d. $N(0,\sigma^2)$ random variables.}
	\item[2.]{We observe the height $h$ at selected locations with spatial frequency $r$ in both directions for a positive integer $r$. All of the observation errors are i.i.d. $N(0,\sigma^2)$ random variables.}
\end{enumerate}
The motivation of using these scenarios is that the he\-ights are in general easier to observe than the velocities (for example, by satellite altimetry).  
In the following experiments, we are going to compare the performance of our proposed 4D-Var method using flow-dependent background covariances with the Hybrid 4D-Var method. In Section \ref{sec:simulationssynthetic} we use sythetic data, while Section \ref{sec:simulationstsunami} we use data from the tsunami waves after the 2011 Japan earthquake.
\subsection{Comparison based on synthetic data}\label{sec:simulationssynthetic}
First we compare the performance of various methods using synthetic data. The shallow water equations were solved on the torus $[0,L]^2$ with $L=210\mathrm{km}$. The  initial condition $\U(0):=(u(0),v(0),h(0))$,
ocean depth $H$ and other ODE parameters were chosen as follows, 
\begin{gather*}
 u(0)=0.5+0.5 \sin( \tfrac{2\pi(x+y)}{L}), \\
 v(0)=0.5-0.5 \cos( \tfrac{2\pi(x-y)}{L}), \\
  h(0)=2\sin(\tfrac{2\pi x}{L})\cos(\tfrac{2\pi y}{L}); \\ 
 H=100+100(1+0.5\sin(\tfrac{2\pi x}{L}))(1+0.5\sin(\tfrac{2\pi y}{L}));\\ 
 \nu=10^{-3},\quad c_{b}=10^{-5},\quad g=9.81, \quad f=10^{-4}. 
\end{gather*}
The discretised versions of the initial condition and the ocean depth were obtained under the choices $d=21$, $\Delta=10 \mathrm{km}$. In the first observation scenario we have chosen the spatial frequency of the velocity observations as $r=3$ (giving $49$ velocity observations). All of the heights are also observed. Observations are made every $10$ seconds, the total observation time is 1 day, and the observation errors had standard deviation $\sigma=10^{-2}$.

In the second observation scenario we have chosen the spatial frequency of the height observations as $r=3$ (giving $49$ height observations). Velocities are not observed. Observations are made every $60$ seconds, the total observation time is 10 days, and the observation errors had standard deviation $\sigma=10^{-2}$.

For the Hybrid 4D-Var, we have used localisation, as described in Section 8.3 of \cite{reichcotterbook}. 
This localisation matrix was chosen as $(C)_{k l}=\rho \l(\frac{r_{l,l}}{r_{\mathrm{loc}}} \r)$, where 
$r_{k,l}$ denotes the spacial distance on the torus between two gridpoints $k$ and $l$, and $\rho$ is the filter function describing the decay of correlations, and $r_{\mathrm{loc}}$ is the localisation radius.
The filter function in the localisation was chosen according to equation (8.29) of \cite{reichcotterbook} as
\begin{align}
\rho(s)=\left\{\begin{matrix} &1-\frac{5}{3}s^2+
\frac{5}{8}s^3+
\frac{1}{2}s^4-\frac{1}{4}s^5 \\
&\text{for $0\le s\le 1$}\\
&-\frac{2}{3}s^{-1}+4-5s+\frac{5}{3}s^2+\frac{5}{8}
s^3-\frac{1}{2}s^4+\frac{1}{12}s^5\\
&\text{for $1\le s\le 2$,}\\
&0  \\ &\text{otherwise.}
\end{matrix}\right.	
\end{align}
We have also used multiplicative ensemble inflation, as described in Section 8.2 of \cite{reichcotterbook}.
This consists of rescaling the ensemble members around their mean by a factor $1+c_{\mathrm{inf}}$ for some $c_{\mathrm{inf}}>0$. Finally, the climatological covariance $\mtx{B}_0$ and mean were estimated from the true unobserved path of the system during the total assimilation time (1 day in our first experiment, and 10 days in our second). This is typically estimated from past data, or by running the model over longer time periods (see \cite{fairbairn2014comparison}), but this was not possible as the non-linear shallow-water equations suffer from numerical instabilities over long time periods (due to the breaking waves phenomenon). The initial ensemble was sampled from a Gaussian distribution corresponding to the estimated climatological mean, and climatological covariance. We have used hybridization so the ensemble based flow-dependent covariances were combined with the climatological covariances according to a hybridization parameter $c_{\mathrm{hyb}} \in [0,1]$.

	Table \ref{tab:hybparameters} states the values of the localization radius, multiplicative inflation, and hybridization parameters that we tested in a grid search for two experiments. In total $3
	\times 3 \times 6  = 54$ different values were tested, with $c_{	\mathrm{hyb}}=0$ corresponding to 4DEnVar.


\begin{table*}[htbp]
\centering
\begin{tabular}{| l | l | l | l |}
	\hline
   Parameter \& Experiment & Synthetic 1 day & Synthetic 10 days & Tsunami \\
  \hline			
  $r_{\mathrm{loc}}$ (localisation) & 2,3,4 & 2,3,4 & 2,3,4\\
  \hline			
  $c_{\mathrm{inf}}$ (inflation)  & $2.5 \cdot 10^{-4}$, $5\cdot 10^{-4}$,  $10^{-3}$ & $5\cdot 10^{-4}$,  $10^{-3}$, $2\cdot 10^{-3}$  & 0.01, 0.02, 0.03 \\
  \hline			
  $c_{\mathrm{hyb}}$ (hybridization) & 0, 0.1, 0.3, 0.5, 0.7, 0.9 &  0, 0.1, 0.3, 0.5, 0.7, 0.9 & 0, 0.1, 0.3, 0.5, 0.7, 0.9 \\
  \hline  
  Ensemble size & 200 & 200 & 100\\
  \hline  
\end{tabular}
  \caption{Tested parameter values for Hybrid 4D-Var}\label{tab:hybparameters}
\end{table*}

\begin{figure}[htbp] 
\begin{center}
\includegraphics[width=8cm]{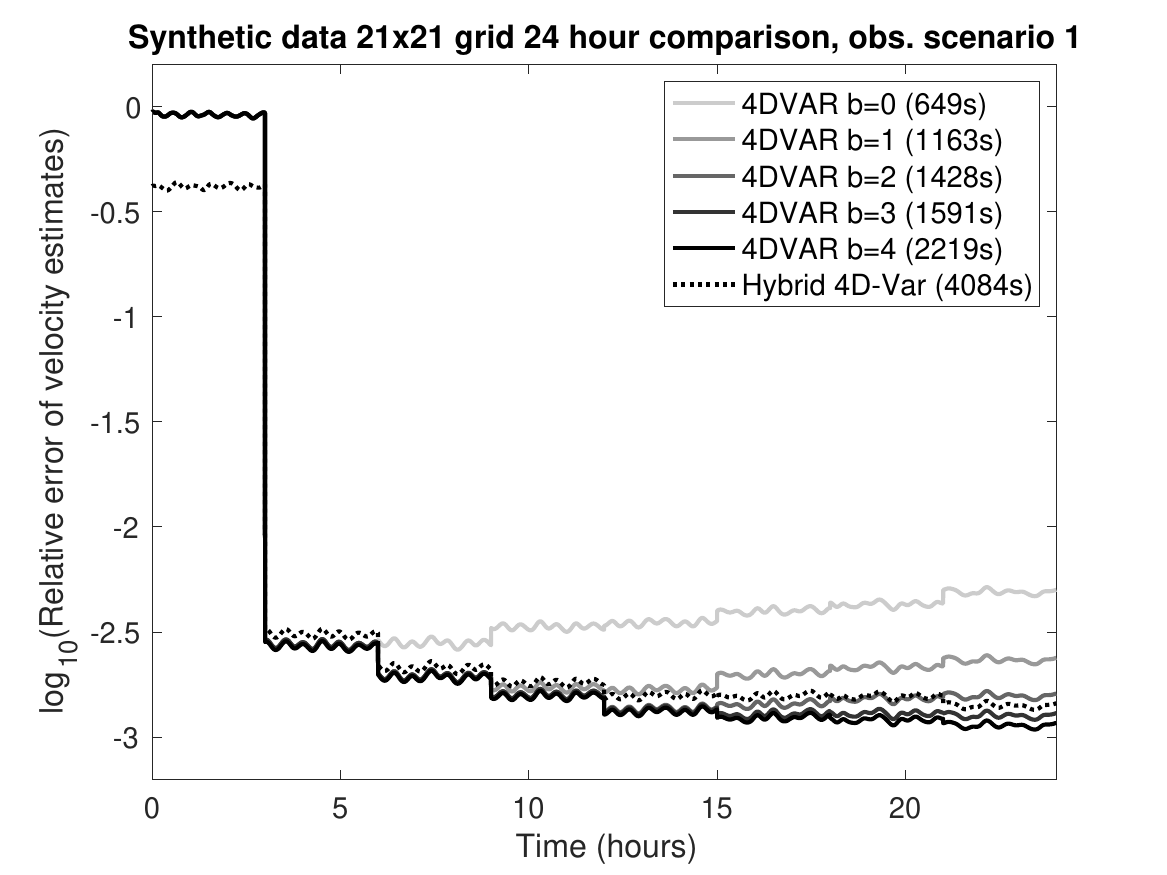}
\end{center}
\vspace{-0.6cm}
\caption{Relative errors of velocity estimates in the case of synthetic data for all methods.
Setting: $d=21$, $k=1080$, $T=3$h, $\sigma=10^{-2}$, $\Delta=10^4$, 10 seconds between observations. The 4D-Var and Hybrid 4D-Var methods process the data is batches of size $k$, hence the filtering accuracy is poor until we have reached the end of the first observation window. The best parameter values for Hybrid 4D-Var were $r_{loc}=4$, $c_{\mathrm{hyb}}=0$,  $c_{\mathrm{inf}}=2.5\cdot 10^{-4}$ (see Table \ref{tab:hybparameters} for all of the parameter values that we tested).}
\label{longrunsc2}
\end{figure}

	In the case of the  proposed 4D-Var method with flow-dependent background covariances, the initial covariance $\mtx{B}_0$ was chosen as a diagonal matrix. The assimilation window $T$ was chosen as 3 hours (which offered the best performance for fixed background covariance). 
	Fig.~\ref{longrunsc2} illustrates the performance of the various methods in the first observation scenario for the 1 day run case. 

 The 4D-Var method was optimised based on the Gauss-Newton method with preconditioned conjugate gradient (PCG) based linear solver. We did not use any preconditioner, and the maximum number of iterations per PCG step was set to 100 (which was sufficient for reducing the relative residual below $0.01$ in most cases). In the Hybrid 4D-Var method, we used a hybrid version of the ENKF based on fixed covariances (i.e. a linear combination of them), and the optimization was done in a similar way as for the 4D-Var method. All of the methods were implemented in Matlab and ran on a single node of the Oxford ARC Arcus-B HPC cluster (16 cores per node). The measure of performance is the relative error of the unobserved component at a certain time $t$, i.e. if $\vct{w}(t)\in \R^{n-n^{\circ}}$ denotes the true value of the unobserved component, and $\hat{\vct{w}}(t)\in \R^{n-n^{\circ}}$ is the estimator, then $\|\hat{\vct{w}}(t)- \vct{w}(t)\|/\|\vct{w}(t)\|$ is the relative error ($\| \cdot \|$ refers to the Euclidean norm).

\begin{figure}[htbp] 
	\begin{center}
	\includegraphics[width=8cm]{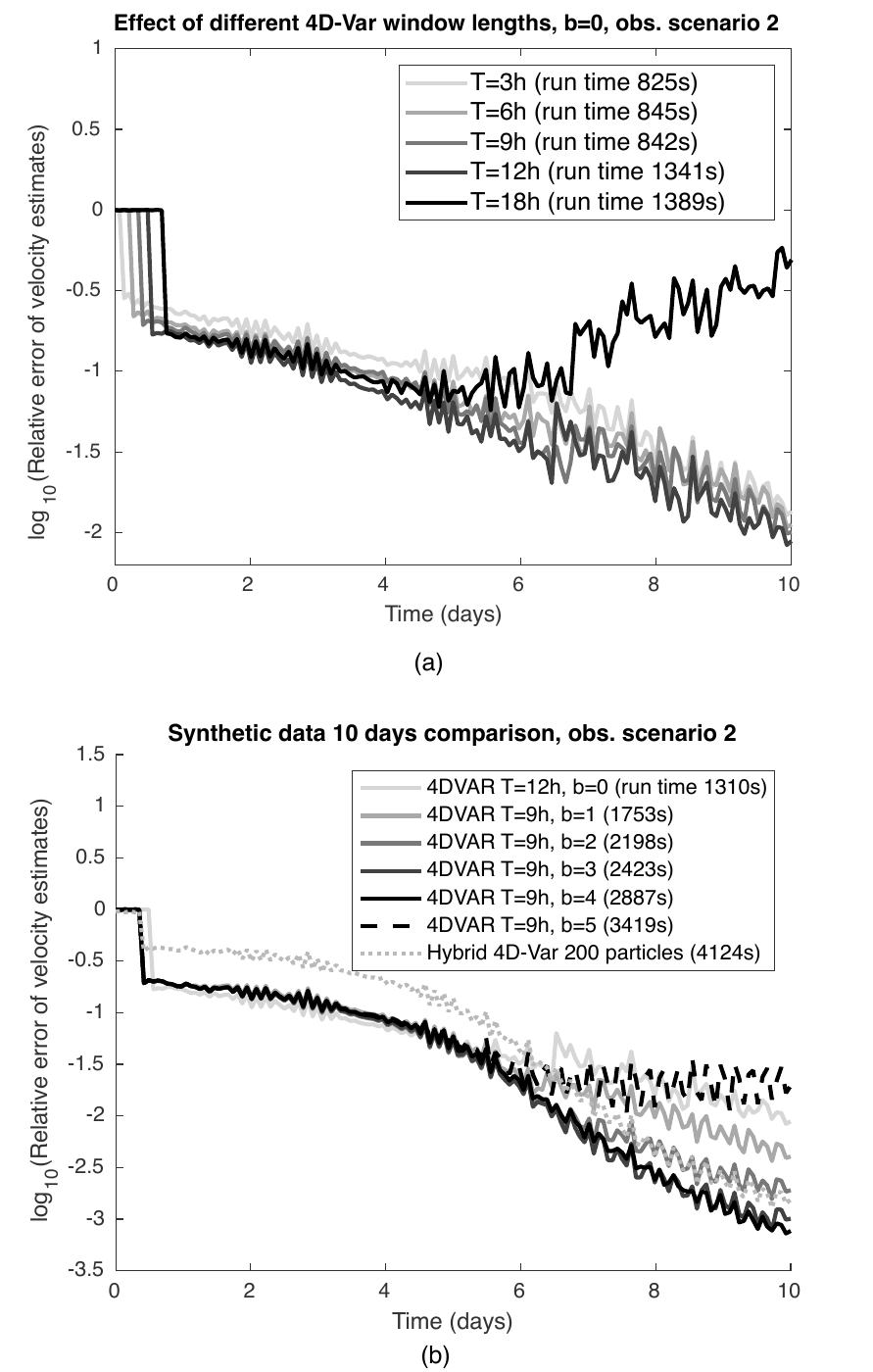}
	\end{center}
	\caption{\label{fig10days2fig} Relative errors of velocity estimates for synthetic data with observation scenario 3.	Setting: $d=21$, $\sigma=10^{-2}$, $\Delta=10^4$, 60 seconds between observations. Fig.~\ref{fig10days2fig}a shows the performance of 4D-Var with fixed covariance matrix for different time lengths. Fig.\ref{fig10days2fig}b compares the performance of our method with Hybrid 4D-Var. The best parameter values for Hybrid 4D-Var were $r_{loc}=2$, $c_{\mathrm{hyb}}=0$,  $c_{\mathrm{inf}}=5\cdot 10^{-4}$ (see Table \ref{tab:hybparameters} for all of the parameter values that we tested) .}
	\label{longrunsc3}
\end{figure}

We have also repeated the experiment in the more challenging second observation scenario. Fig.~\ref{fig10days2fig}a shows the performance of 4D-Var with a fixed background covariance matrix with varying window sizes $T=3\mr{h}$, $6\mr{h}$, $9\mr{h}$, $12\mr{h}$ and $18\mr{h}$. In the case of $12\mr{h}$ and $18\mr{h}$, we have used the idea of \cite{pires1996extending} to first find the optimum for shorter windows, and then gradually extend the window length to $T$ to avoid issues with non-convexity. This has resulted in better optimum at the cost of longer computational time (it did not make a difference at shorter window lengths). Overally, we can see that the $T=12\mr{h}$ has the best performance, but the computational time is longer than for $T=9\mr{h}$ (as we in fact first find the optimum based on the first half of the observations in the window, and then continue with the other half). At $T=18h$ the performance diminishes due to the non-convexity of the likelihood, and even the gradual extension of the window length fails to overcome this problem.

Fig.~\ref{fig10days2fig}b compares the performance of our method (based on $T=9h$, but with choices of $b$ from 1 to 5) with 4D-Var with fixed window length ($T=12h$), ENKF and Hybrid 4D-Var. For this synthetic dataset, our 4D-Var-based method with $b=3$ offered the best performance. $b=4$ was similar but with higher computational cost, and $b=5$ resulted in worse performance, likely due to the non-linearity of the system. We can see that using observations in earlier assimilation windows to update the background covariance matrix in a flow-dependent way is very beneficial, with relative errors reduced by as much as 70-90\% compared to using a fixed background matrix.

\begin{figure}[htbp] 
	\begin{center}
	\includegraphics[width=8cm]{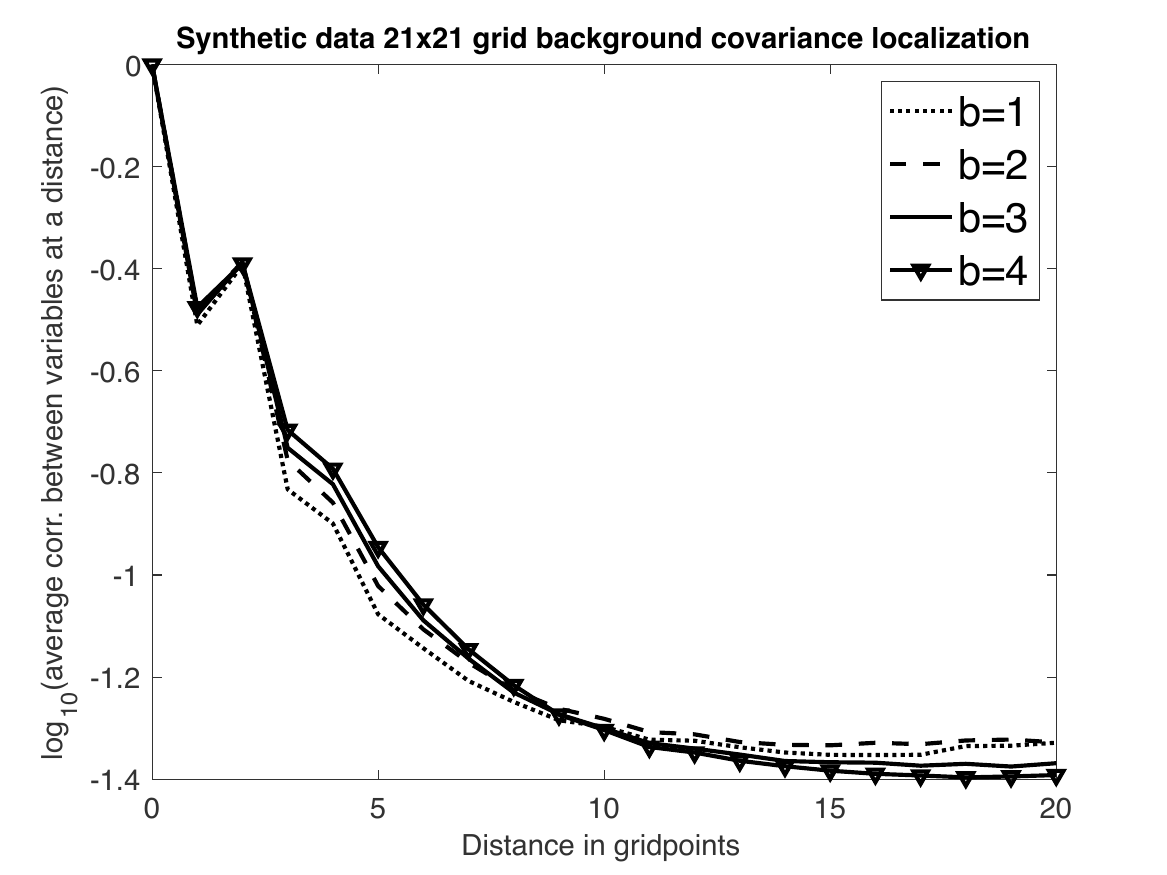}
	\end{center}
	\vspace{-0.6cm}
	\caption{Correlation decay in flow-dependent background covariances for different values of $b$. Setting: $d=21$, $k=540$, $T=9h$, $\sigma=10^{-2}$, $\Delta=10^4$, observation scenario 2.}
	\label{figcovloc}
\end{figure}

To better understand the reason for this improvement in performance, on Fig.~\ref{figcovloc} we have plotted the average correlations in background covariances between the components at a given distance, in the cases $b=1, 2, 3$, for the first observation scenario computed at the last observation window (after 24 hours). As we can see, as $b$ increases, the background covariance matrix changes and becomes less-and-less localised.

The performance of the Hybrid 4D-Var was quite good and it considerably improved upon using a fixed background covariance matrix, but nevertheless our met\-hod still had significantly better accuracy, especially in second observation scenario which involved data assimilation over a longer time period (10 days) with less frequent observations. We believe that this increase in accuracy is due to the more accurate modelling of background covariances, which become less-and-less localised over longer time periods. 



\subsection{Comparison based on tsunami data}\label{sec:simulationstsunami}
\begin{figure}[htbp] 
\begin{center}
\includegraphics[width=8.5cm]{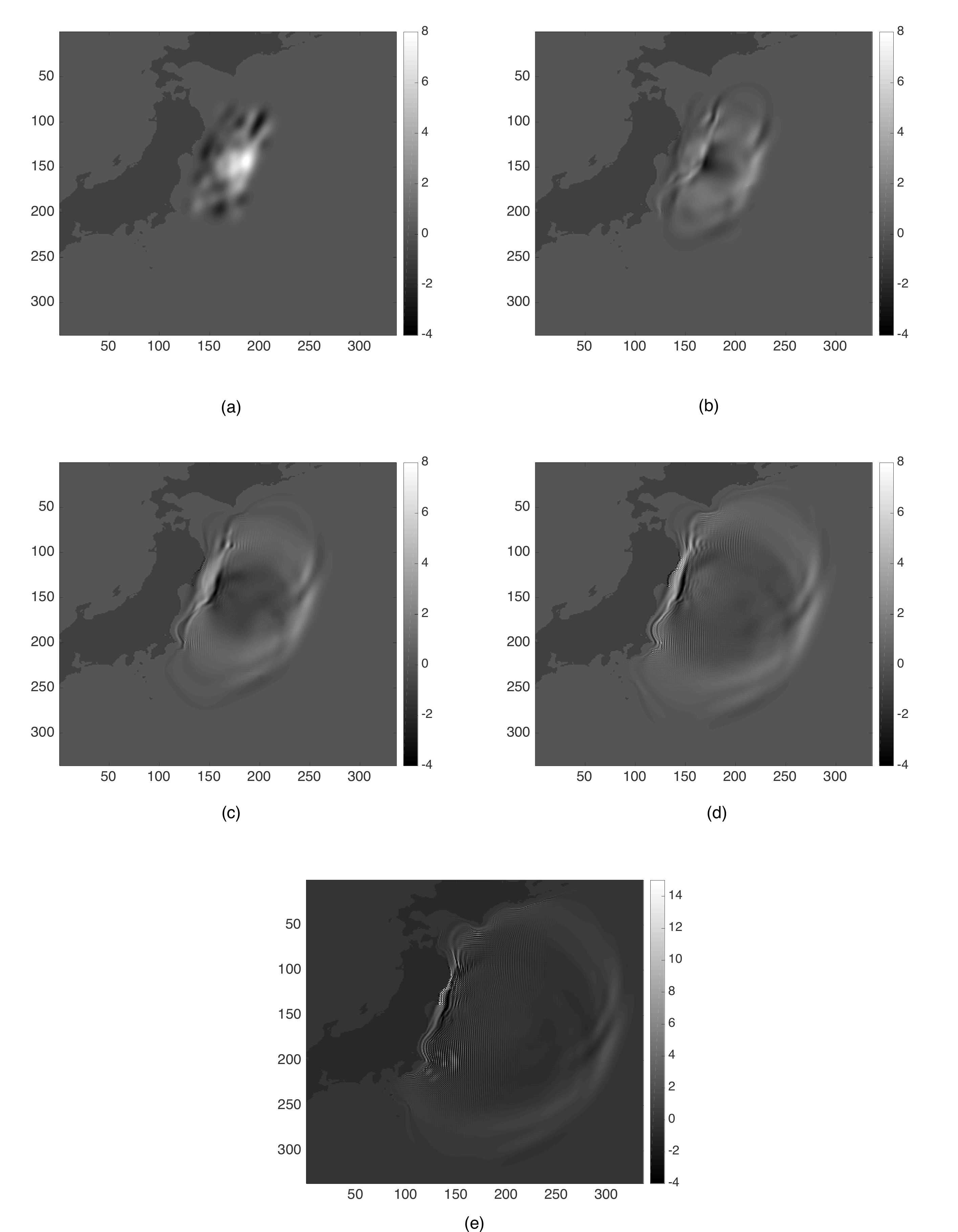}
\end{center}
\caption{\label{tsusimfig}Evolution of the height of the tsunami waves (in meters) at 0, 10, 20, 30, and 40 mins (for grid size $d=336$).}
\end{figure}

The shallow water equations are applied in tsunami modelling. \cite{saito2011tsunami}  estimate the initial distribution of the tsunami waves after the 2011 Japan earthquake. They use  data from 17 locations in the ocean, where the wave heights were observed continuously in time. We have used these estimates as our initial condition for the heights, and set the initial velocities to zero (as they are unknown). Using publicly available bathymetry data for $\ol{h}$, and the above described initial condition, we have run a simulation of 40 minutes for our model, see Fig.~\ref{tsusimfig}. We have tested the efficiency of the data assimilation methods also on this simulated dataset, considering a time interval from 10 to 40 minutes (thus the initial condition corresponds to the value of the model after 10 minutes and is shown in Fig.~\ref{tsusimfig}b). Due to the somewhat rough nature of the tsunami waves, in this example we have found that setting the background precision (inverse covariance) matrix $\B_0^{-1}$ as zero offered the best performance for the proposed 4D-Var method, while we used a diagonal matrix for the Hybrid 4D-Var method. The localisation and ensemble inflation was implemented as described in Section \ref{sec:simulationssynthetic}, with the tested parameter values shown in Table \ref{tab:hybparameters}. 
The 4D-Var method was optimised based on the Gauss-Newton method with preconditioned conjugate gradient (PCG) based linear solver without any preconditioner, and the maximum number of iterations per PCG step was set to 500.

\begin{figure}[htbp] 
\begin{center}
\includegraphics[width=8cm]{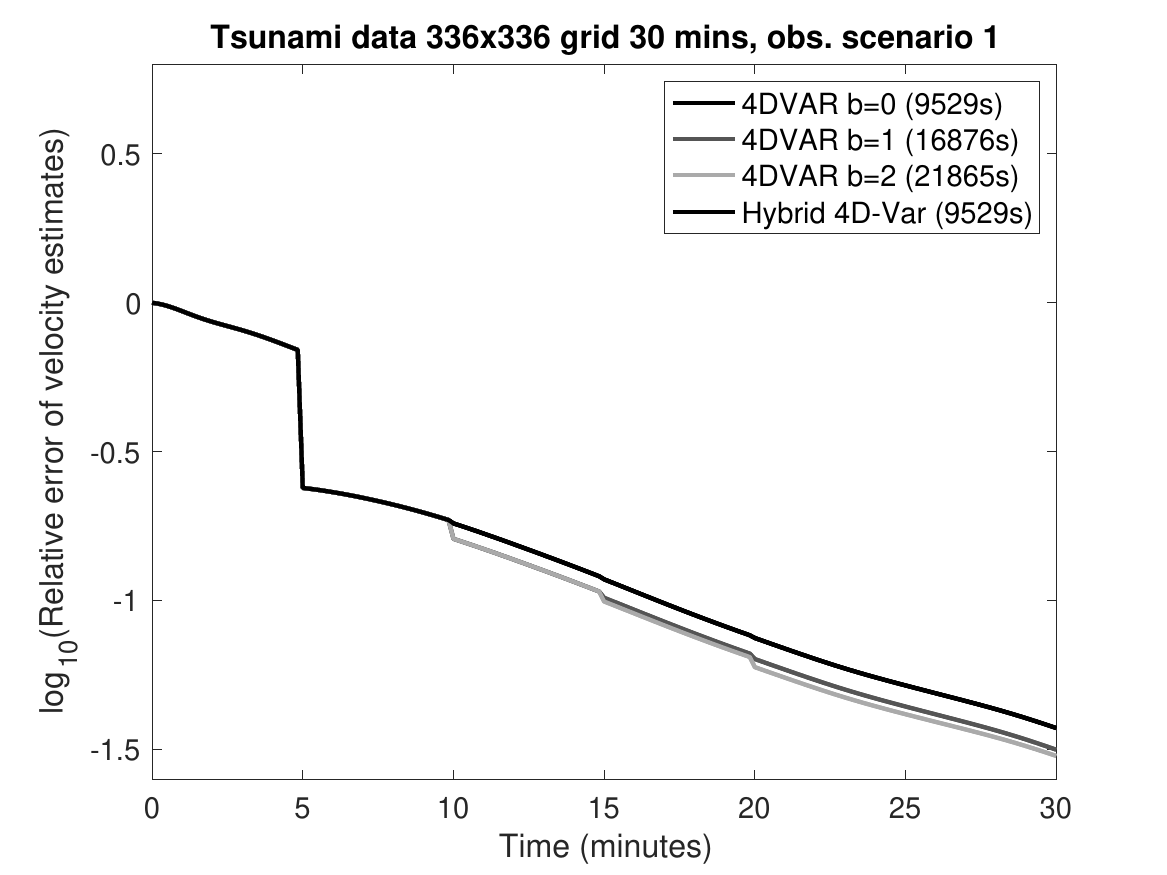}
\end{center}
\caption{Relative error of estimates of velocities for tsunami data in the first observation scenario, all methods. Setting: $d=336$, $k=30$, $T=30$mins, $\sigma=10^{-2}$. The best performance for hybrid Hybrid 4D-Var was obtained by only using the fixed background covariance matrix, hence the performance and running time is equivalent to 4D-Var with $b=0$ (see Table \ref{tab:hybparameters} for all of the parameter values that we tested).}
\label{tsunamifig}
\end{figure}

Fig.~\ref{tsunamifig} compares the performance of the methods for this synthetic dataset implemented for grid size $d=336$ (so the dimension on the dynamical system is $n=3d^2=338,688$) in the first observation scenario, where the spatial frequency of the velocity observations was chosen as $r=48$ (i.e. $7\cdot 7=49$ velocity observations in total).

As in the previous synthetic example, the proposed 4D-Var based method offer the best performance. Note that the best performance of the Hybrid 4D-Var was achieved when we have used only a fixed background covariance matrix, and coefficient for the ENKF based background covariance is set to zero in the hybridization (see Table \ref{tab:hybparameters} for the tested parameter values for hybridization, inflation and localization). Hence in this complex highly unstable situation the ENKF based background covariances did not help, while our proposed flow-dependent covariances improved the precision of the velocity estimates when using $b=1$ and $b=2$.

\section{Conclusion}\label{secconclusion}
In this work we have presented a new method for updating the background covariances in 4D-Var filtering, and applied it to the shallow-water equations. Our method finds the MAP estimator of the initial position using the Gauss-Newton's method with the Hessian matrix stored and the background covariances obtained in a factorised form. Our method is computationally efficient and has memory and computational costs that scale nearly linearly with the size of the grid. 
        
4D-Var-based methods are less directly parallelisable compared to ENKF as the optimization steps and the ODE solver steps are indeed inherently serial. Hence in a parallel environment, it is likely that the background covariance part of the Hybrid 4D-Var can be computed significantly faster using the ENKF based methods compared to the proposed method. However, we have shown in the experiments that the method proposed by this paper can be significantly more accurate in the perfect model scenario for the shallow-water equations. Moreover, both Hybrid 4D-Var and the proposed method use the same computations based on adjoint equations and tangent linear model for the data in the current assimilation window (they only differ in formulation of the flow-dependent background covariances). The total computational time of the background covariances takes typically at most factor of $b$ times longer than the computations for the data in the current window, thus the proposed method is not overly computationally expensive.

It remains to be seen if the improvements in accuracy for the proposed method also hold in more complex weather forecasting models, in the presence of some model error. Our hope is that the proposed method could yield significant improvements in accuracy in some challenging data assimilation scenarios where modelling covariance localization is difficult, without the need of extensive tuning.

The Matlab code for the experiments is available at \url{https://github.com/paulindani/shallowwater}.
 
\subsubsection*{Acknowledgements}
The authors thank Joe Wallwork for providing us the tsunami data set, and for our correspondence related to the shallow-water equations. 
All authors were supported by an AcRF tier 2 grant: R-155-000-161-112. AJ was also supported by KAUST baseline funding. This material is based upon work supported in part by the U.S. Army Research Laboratory and the U.S. Army Research Office, and by the U.K. Ministry of Defence (MoD) and the U.K. Engineering and Physical Research Council (EPSRC) under grant number EP/R013616/1. DC was partially supported by the EPSRC grant: EP/N023781/1. AB was supported by a Leverhulme Trust Prize.
\appendix
\section{Sparse storage of Jacobians for Shallow-Water Dynamics}
In this section we explain a possible method for the computation and storage of the Jacobians $\M_i$, $1\le i\le k$ specifically for the case of the shallow-water equations \eqref{ueq}-\eqref{heq}. For other equations, it might be the case that storing $(\M_i)_{1\le i\le k}$ directly as follows is not practical because the interaction between the components is not local and the Jacobian matrix is not sparse. In such cases, we can still apply the tangent-linear and adjoint equations for computing the matrix-vector products $\M_i v$ and $\M_i^{T} v$, as explained in 
Section \ref{secflowdependent4DVARfilter}.

One can observe that time derivatives at a grid position only depends on its grid neighbours. Moreover, the shallow-water equations are of the general form $\frac{d\x}{dt}=-\mtx{A}\x-\mtx{B}(\x,\x)+\vct{f}$, where $\mtx{A}$ is an $n\times n$ matrix, $\mtx{B}$ is a $n\times n\times n$ array, and $\vct{f}$ is a constant vector in $\R^{n}$ (note that for the shallow-water equations \eqref{ueq}-\eqref{heq}, we have $\vct{f}=0$). For equations of this form, there is an efficient way of calculating the time derivatives and their Jacobians, stated in equations (3.14) and (3.16) of \cite{optimization2017}.
Based on these, one can use Taylor's expansion to compute the Jacobian $\M(t,s)[\x(s)]$, that is
\begin{equation}\label{TaylorJacobianeq}\M(t,s)[\x(s)]\approx \mtx{I}_{n}+\sum_{l=1}^{l_{\max}}\frac{\partial \l(\l.\frac{d^l}{dt^l} M(t,s)[\x(s)]\r|_{t=s}\r)}{\partial \x(s)}\cdot \frac{(t-s)^l}{l!},
\end{equation}
for some $l_{\max}>0$. Due to the fact that the first derivatives only contain terms from neighbouring gridpoints, it is easy to see that the above approximation only has non-zero elements for gridpoints that are no more that $l_{\max}$ steps away. This means that as long as $t-s$ is sufficiently small, the Jacobian $\M(t,s)[\x(s)]$ can be stored as a sparse matrix with $\mathcal{O}(n)$ non-zero elements. If the time interval between the observations is sufficiently small, then  each of $\M_1,\ldots, \M_k$ can be stored as a single sparse matrix defined by  \eqref{TaylorJacobianeq}. The inverse of the Jacobian satisfies that $(\M(t,s)[\x(s)])^{-1}=\M(s,t)[\x(t)]$, so it can be calculated by  \eqref{TaylorJacobianeq} with terms $(s-t)^l$ instead of $(t-s)^l$ and $\x(t)$ instead of $\x(s)$.

At this point we note that one could attempt to use the Jacobians $\M(t_l,t_0)$ directly. However, for $l\ll n$, storing the Jacobians $\M_1,\cdots, \M_l$ separately requires $\mathcal{O}(n l)$ memory, and the effect of $\M_l \M_{l-1}\cdots \M_{1}$ on a vector can be evaluated in $\mathcal{O}(n l)$ time, while for 2D lattices, the product $\M_l\cdots \M_1$ would require $\mathcal{O}(nl^2)$ memory, and its effect on a vector would require $\mathcal{O}(n l^2)$ time to evaluate (for 3D lattices, it would incur up to $\mathcal{O}(n l^3)$ memory and computational cost).
For the same reason, for longer time intervals between observations, it is more effective to break the interval  into $r>1$ smaller blocks of  equal size, and store the Jacobians corresponding to each of them. In this case, when applying the Jacobian $\M_l$ on a vector, the result can computed as the product of the Jacobians for the shorter intervals.

\bibliographystyle{spbasic}
\bibliography{References}

\end{document}